\documentclass[]{spie}  

\usepackage{amsmath,amsfonts,amssymb}
\usepackage{graphicx}
\usepackage[colorlinks=true, allcolors=blue]{hyperref}

\title{A genetic algorithm for peer-review panel composition}

\author[a]{Ferdinando Patat}
\affil[a]{European Southern Observatory, Karl Schwarzschildstr. 2, D-85748, Garching b. M\"unchen, Germany}

\authorinfo{Further author information: (Send correspondence to F. Patat)\\E-mail: fpatat@eso.org}

\begin{document} 
\maketitle

\begin{abstract}
The composition of scientific review panels is a constrained optimization problem in which a finite pool of experts must be distributed among multiple panels while balancing scientific expertise and demographic diversity. As the number of possible panel configurations grows very rapidly with the number of reviewers, exhaustive searches rapidly become computationally impractical.
In this paper I present a genetic algorithm designed to optimize panel composition for the European Southern Observatory (ESO) proposal evaluation process. Panel assignments are represented through a chromosome-based encoding. Candidate solutions are evaluated using a fitness function based on four imbalance indicators: scientific expertise, gender, country affiliation, and professional seniority.
The method is tested using real reviewer data from the ESO proposal handling system. The results show that the genetic algorithm rapidly identifies panel configurations with substantially lower imbalance than those obtained from random assignments and progressively improves the quality of the overall population. Beyond producing a single optimized configuration, the approach generates a set of high-quality panel realizations that can subsequently be filtered according to additional operational constraints not explicitly included in the fitness function. Although developed for ESO, the methodology is general and applicable to a wide range of panel-based peer-review systems.
\end{abstract}

\keywords{Telescope time allocation, Peer review systems, Observatory management, Decision support systems, Optimization algorithms}

\section{INTRODUCTION}
\label{sec:intro}

With small variations on the same theme, basically all major astronomical facilities collecting a large number of proposals adopt a common review schema. This consists in first classifying the submitted proposals in scientific categories and then assigning them to panels, typically several per category, depending on the categorization schema. Because of the logistics of the process, the reviewers need to be recruited well ahead of the proposal submission deadline. This implies that the facilities recruit the reviewers, both in terms of number and expertise, based on the statistics of the previous calls and the needs dictated by the expertise areas left vacant by leaving members. One would be tempted to think that the order should be inverted: one should first receive the proposals and only then select the reviewers who best match the specific scientific areas. Even if this were at all possible, it would nevertheless not solve the problem, as the number of reviewers is forcedly limited, an aspect which necessarily restricts the possibility of finding a sufficiently large number of experts for any received proposal. The composition of ad hoc panels is only possible with the Distributed Peer Review (DPR), in which one has the luxury of being able to practically create one panel per proposal.\cite{Patat2019,Kerzendorf2020, AlmaDPR} This is because DPR is a self-sustaining process: the number of reviewers linearly grows with the number of proposals. For the classical panels this is obviously not the case. In addition, the basic principle behind the panel concept is that all members review the same set of proposals. This is the fundamental requirement which enables the panel discussion and the formulation of a collegial evaluation, but it also introduces quite some rigidity in the process.

Having recruited the panel members is already a significant achievement. In an epoch in which several facilities attract a very large number of proposals, peer review comes with a high price tag for the community. Serving on a panel typically adds on top of other duties not strictly related to pure scientific research activities, such as reviewing papers, serving on boards, teaching, and supervising students. In addition, the most desirable reviewers are often those in greatest demand. As a consequence, assembling a suitably balanced set of reviewers has become increasingly challenging.
Several efforts have been made to automate different aspects of the proposal review process. For example, Strolger et al.\cite{Strolger17} developed the PACMan system for the HST and JWST review process, using machine-learning techniques to assist proposal categorization, panelist selection, and reviewer assignment. These approaches aim at improving the match between proposals and reviewers and reducing the operational burden associated with large review programs. In fact, systems such as PACMan and the method presented here are largely complementary: the former focuses on proposal categorization and reviewer matching, while the latter focuses on the composition of balanced review panels.

The problem addressed in this paper is different. Assuming that a pool of reviewers has already been recruited and assigned to a broad scientific category, the objective is to determine how these reviewers should be distributed among the panels of that category in order to maximize expertise coverage and diversity while minimizing a set of predefined imbalance indicators. To this end, I present a simple genetic algorithm that optimizes panel composition through the minimization of a weighted fitness function.

\section{PANEL HANDLING AT ESO}
\label{sec:eso}

In recent years, the ESO proposal review process has evolved significantly. A major development has been the introduction of Distributed Peer Review (DPR), which was designed to alleviate the increasing workload of the traditional review panels by transferring a substantial fraction of proposal assessments to this new review channel.\cite{Patat2019,Kerzendorf2020,Jerabkova2023,Jerabkova2025} However, the panel-based component of the process has remained largely unchanged and is described in detail by Patat \& Hussain.\cite{PatatHussain2013} Only the aspects relevant to the present work are summarized here.

ESO proposals are classified into four broad scientific categories. Depending on proposal pressure, each category is divided into panels, whose members evaluate the same set of proposals and subsequently meet to establish a collective ranking.
Panel members are recruited from the international astronomical community before proposal submission.\footnote{The recruitment is based on the nominations of the Users Committee and formally approved by the Nominating Committee.} As a consequence, panel composition must be based on anticipated scientific demand rather than on the actual proposals received. Once the reviewer pool for a given category has been assembled, the reviewers must be distributed among the corresponding panels. The goal is to ensure that all panels provide comparable coverage of the relevant scientific expertise while avoiding significant imbalances in demographic characteristics such as gender, geographical affiliation, and seniority. Additional constraints may also apply: all panels should contain the same number of members to guarantee a fair and homogeneous evaluation of proposals, and specific roles, such as panel chairs, need to be assigned in advance.

Each reviewer is characterized by a unique identifier, country of affiliation, professional seniority, gender, and a set of scientific keywords ranked by decreasing relevance.\footnote{In the current ESO review implementation, reviewers define these keywords themselves and rank them according to their expertise.}
In addition, each reviewer is assigned to a scientific category during the recruitment process. Consequently, all reviewers within a given category possess at least one keyword relevant to that category. In the following, we consider the panels associated with a single scientific category, treating each category independently. Under this assumption, every reviewer belongs to one and only one category. Although reviewers may indicate their suitability for multiple categories, the final assignment is determined during recruitment according to the overall requirements of the review process.

While individual implementations differ in detail, most major astronomical facilities adopt broadly similar panel-based review systems. The optimization problem discussed in this paper is therefore not unique to ESO, but is common to many proposal evaluation processes.

\section{DEFINITIONS AND FORMALISM}
\label{sec:def}

The composition of the panel can be formulated as an optimization problem, in which one or more figures of merit (FoMs) are to be maximized. Once these FoMs are defined, the task becomes evaluating the possible panel configurations, computing their FoMs, and selecting the optimal solution. In principle, a brute-force approach could be employed, generating all possible panel configurations, evaluating each one, and selecting the best. However, it is readily apparent that such an approach would be computationally prohibitive.
Let us consider a set of $N$ experts to be distributed in $M$ panels, each containing $n_l$ members ($l=1,2,\ldots,M$), obeying to the condition $\sum_{(l=1)}^M n_l =N$. The number of possible panel realizations is equal to the number of ways to partition $N$ experts into M non-ordered groups (the panels) where the order of elements within the groups does not matter, but the sizes of the panels are fixed ($n_1,n_2,\ldots,n_M$). This is a combinatorial partitioning problem known as multi-set partitioning,\cite{Aigner1997} where the number of possible configurations is given by:

\begin{displaymath}
P_{(n_1, n_2, \ldots, n_M)} = \frac{N!}{n_1! \, n_2! \, \ldots \, n_M!}
\end{displaymath}

As this is a typical configuration, in the following we will assume that all panels have the same size $n=N/M$ ($n_1=n_2=\ldots=n_M$). For instance in the case of $N=24$ members to be distributed in $M=4$ panels with $n=6$ members each, the number of partitions is:\footnote{This is not completely correct as, in the practical case, once the panel chairs are selected (which is typically done manually, based on several considerations), only the remaining panel members can be moved across the panels within their scientific categories.}

\[
P_{6,6,6,6} = \frac{24!}{(6!)^4} \approx 2.3 \times 10^{12}
\]

This shows that a brute-force approach, though virtually possible, is numerically extremely expensive. For this reason, optimization algorithms need to be adopted. Among these, we consider here a Genetic Algorithm (GA), a population-based optimization technique inspired by biological evolution and widely used in combinatorial optimization problems.\cite{Goldberg89} GAs are particularly effective when the solution space is large and complex, and one needs to explore a wide variety of potential solutions.

\begin{figure} [t]
\begin{center}
\begin{tabular}{c}
\includegraphics[width=\linewidth]{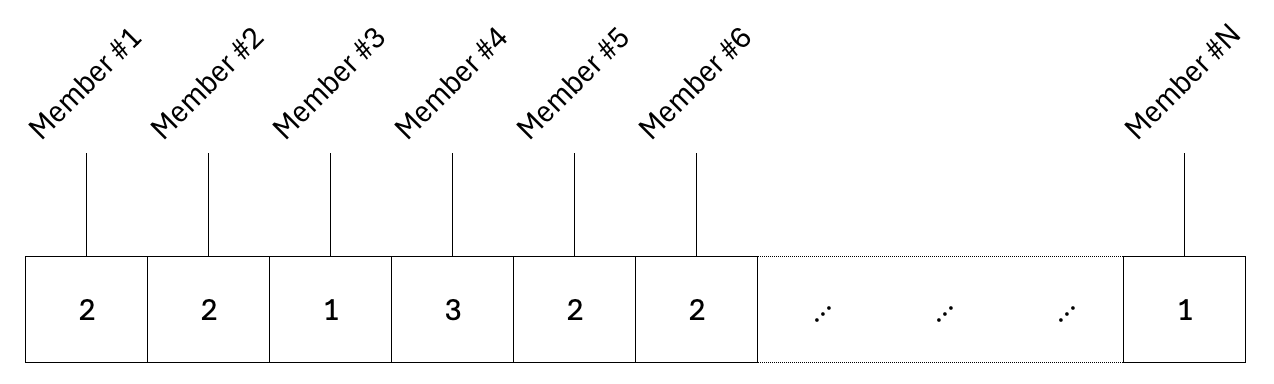}
\end{tabular}
\end{center}
\caption{\label{fig:chromosome} 
Example index representation of a chromosome specifying a panel composition within a category with 3 panels (1, 2, 3) and $N$ total members.}
\end{figure}

In this approach each configuration is represented by a chromosome.
This is essentially a data structure (typically a vector) that encodes the solution's variables. Just like chromosomes in biology carry genes, the chromosome in a GA carries the information that defines a specific solution. The elements of the chromosome vector are called genes. In the following I will adopt the so-called index-based representation for the chromosome.\cite{Falkenauer1998} In this representation, the $i$-th element of the chromosome represents the panel assignment of the $i$-th expert. For example, if there are 3 panels with 6 members each, the chromosome will have 3×6=18 genes, and each gene will be an integer between 1 and 3, indicating which panel the expert is assigned to. An example of this representation is given in Figure~\ref{fig:chromosome}.

With this schema a given panel composition has a unique representation. Different panel compositions are achieved by introducing chromosome mutations which must obey to the constraint that the number of members per panel, the panel size, is fixed. This implies that the number of genes with the same value has to be the same for all the values. This can be achieved implementing mutation operators which either respect this constraint or repair deviating cases.
For instance, one can start from the following initial chromosome:

\[
[1, \mathbf{1}, 1, 1, 1, 1,\; \mathbf{2}, 2, 2, 2, 2, 2,\; 3, 3, 3, 3, 3, 3]
\]

and mutate it by swapping the assignations of members 2 and 7 (indicated in bold face for clarity):

\[
[1, \mathbf{2}, 1, 1, 1, \mathbf{1}, 1, 2, 2, 2, 2, 2,\; 3, 3, 3, 3, 3, 3]
\]

The permutation of two genes implicitly obeys to the fixed panel size constraint. However, swapping two genes with the same value does not mutate the chromosome. To avoid wasting computing time, gene swapping is applied only to genes with different values and only unique permutations are considered.
Another aspect that needs to be taken into account is that more complex genetic processes do not necessarily conserve the panel size. An illustrative example is that of the crossover.
In a genetic algorithm, crossover is the process through which two parent solutions (chromosomes) combine to produce new offspring that inherit characteristics from both parents. The operation aims to explore new regions of the solution space by mixing existing panel assignments in different ways. Typically, a random crossover point (or multiple points) is selected along the chromosome, and the segments before and after this point are exchanged between the two parents to generate two new individuals. This mechanism preserves part of each parent’s structure while introducing variations that can lead to improved solutions. However, when chromosomes represent panel assignments, a simple crossover does not necessarily preserve the correct number of members per panel. This is illustrated by the following example. Let the two parent chromosomes be:

\[
\begin{aligned}
P_1 &= [1,1,1,1,1,1,2,2|,2,2,2,2,3,3,3,3,3,3] \\
P_2 &= [3,3,3,3,3,3,1,1|,1,1,1,1,2,2,2,2,2,2]
\end{aligned}
\]

Suppose we chose a crossover point after the 8th gene (for clarity marked by a $|$). The resulting offspring are:

\[
\begin{aligned}
O_1 &= [1,1,1,1,1,1,2,2|,1,1,1,1,2,2,2,2,2,2] \\
O_2 &= [3,3,3,3,3,3,1,1|,2,2,2,2,3,3,3,3,3,3]
\end{aligned}
\]

The offspring are no longer balanced: $O_1$ has too many members assigned to panels 1 and 2 and none to panel 3. $O_2$ has most reviewers assigned to panel 3, and too few to panel 1 and 2. Such inconsistencies are common when applying crossover operations to constrained combinatorial representations, such as panel assignments. They can be addressed either by applying a repair procedure to restore feasibility after crossover, or by designing a constraint-preserving operator that enforces validity during the generation of offspring.
One example is the guided uniform crossover strategy adopted in this work, inspired by constraint-preserving crossover operators commonly used in grouping problems.\cite{Falkenauer1998}
In this procedure, each child inherits panel assignments from the two parents with equal probability at each gene position, following a standard uniform crossover scheme. To ensure that panel size constraints are satisfied, a quota system is applied such that each panel appears exactly $n$ times in the resulting chromosome. After the initial uniform assignment, any excess or missing allocations are adjusted by reassigning reviewers according to the remaining available panel quotas.
This guided approach effectively combines the genetic diversity introduced by uniform crossover with a deterministic, constraint-preserving mechanism, ensuring that every offspring chromosome remains a valid panel configuration without the need for a separate repair step.

\begin{figure} [t]
\begin{center}
\begin{tabular}{c}
\includegraphics[width=\linewidth]{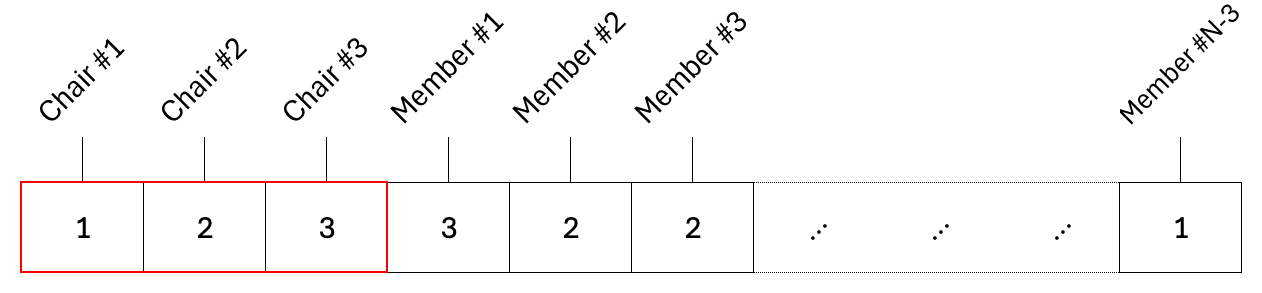}
\end{tabular}
\end{center}
\caption{\label{fig:panelchairs} 
Chromosome representation including the panel chairs in the initial segment, which is not subject to mutations.}
\end{figure}

A final consideration concerns the panel chairs, a role commonly present in review panels. Since chairs are statically assigned to specific panels, their corresponding genes remain fixed and are not subject to mutation. This is implemented by allocating their assignments to the first $M$ positions of the chromosome, which are kept constant throughout the evolutionary process, while only the remaining $N-M$ genes are allowed to mutate. The overall structure of this approach is illustrated schematically in Figure~\ref{fig:panelchairs}. Although only the mutable genes participate in the genetic operations, the entire chromosome will be used in the FoMs evaluation, ensuring that both fixed and variable assignments contribute to the assessment of each individual.

The principle of a GA is to start from a random population of individuals and let them evolve through successive generations.\cite{Goldberg89} Pairs of individuals are selected to reproduce, generating offspring that inherit part of their genetic properties after undergoing genetic operations such as mutation, crossover, and elitism. The new population then reproduces again and the process is repeated. All this is regulated by a fitness function, which describes how well a particular chromosome (a panel composition in our case) satisfies the optimization requirements and implements the principle of survival of the fittest. For each generation the fitness of the best individual is evaluated, and the process stops when no significant improvement is observed. At that point the best individual is taken as the optimal solution. In summary, the GA starts from a random population and, through repeated cycles of reproduction and selection, progressively favors fitter individuals until a near-optimal solution is obtained.

\begin{figure} [t]
\begin{center}
\begin{tabular}{c}
\includegraphics[width=\linewidth]{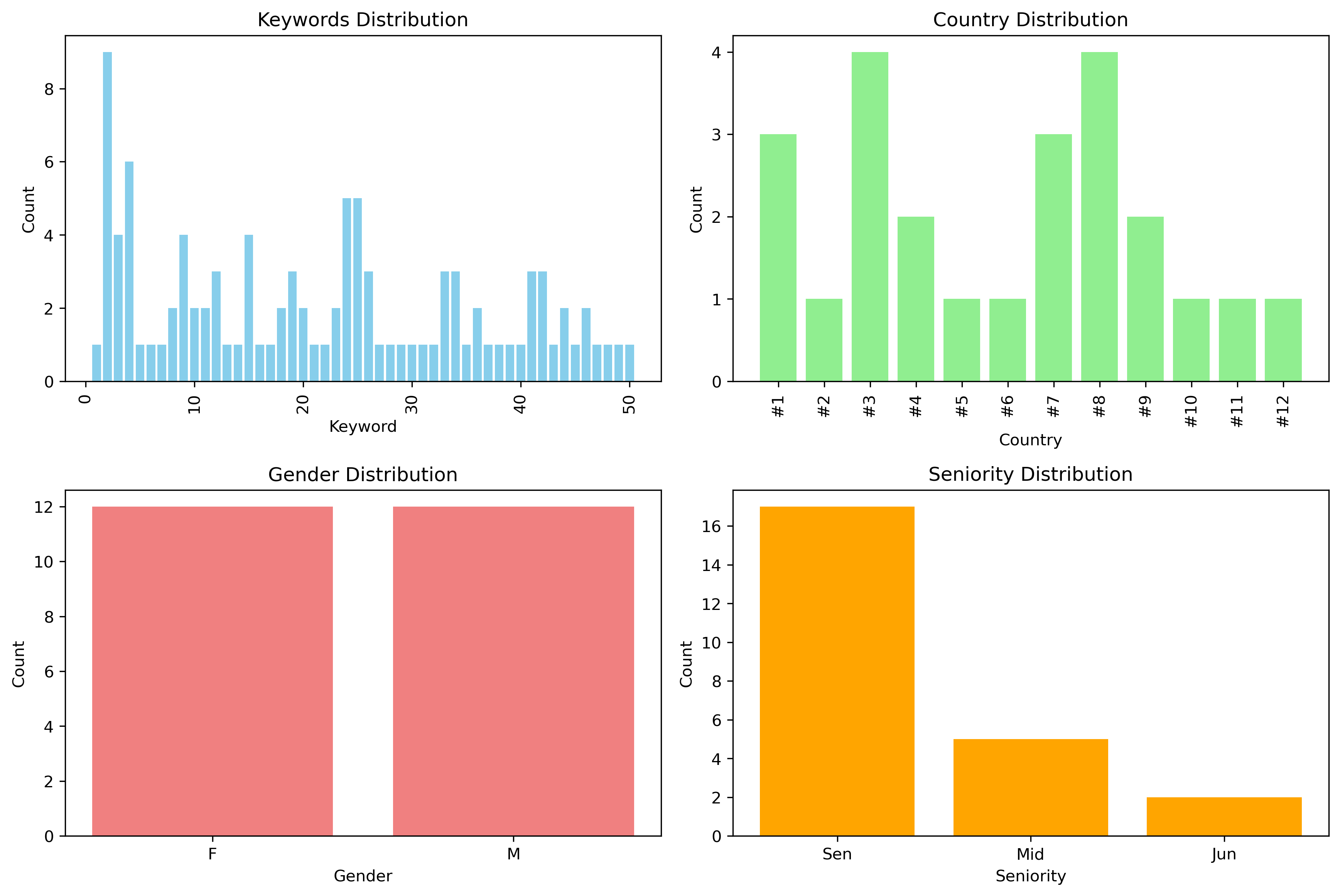}
\end{tabular}
\end{center}
\caption{\label{fig:reviewers} 
Example of real expert properties (ESO D category, Period 117): keywords (upper left), country (upper right), gender (lower left) and seniority (lower right).}
\end{figure} 

\section{FITNESS FUNCTION}

A core component of the method described in the previous section is the definition of the fitness function, which determines how panel compositions are evaluated and optimized. The specific formulation of this function depends on the desired balance between expertise uniformity and specialization across panels.
For illustration, consider two contrasting approaches:

\begin{itemize}
\item {\bf Case A}: all panels exhibit a similar level of expertise within the given category (broad and balanced).
\item {\bf Case B}: each panel is highly specialized, concentrating expertise in a narrow domain (deep but uneven).
\end{itemize}

The specialization of individual reviewers is defined by their sets of keywords, representing their scientific areas of expertise.
In Case A, the fitness function would therefore penalize keyword concentration, while in Case B it would instead reward it.
In the following, we describe the approach adopted at ESO, which prioritizes diversity within panels. This is achieved by minimizing a set of quantities collectively referred to as imbalances, computed across four reviewer attributes: scientific keywords, gender, country of affiliation, and scientific seniority.

\begin{figure} [t]
\begin{center}
\begin{tabular}{c}
\includegraphics[width=\linewidth]{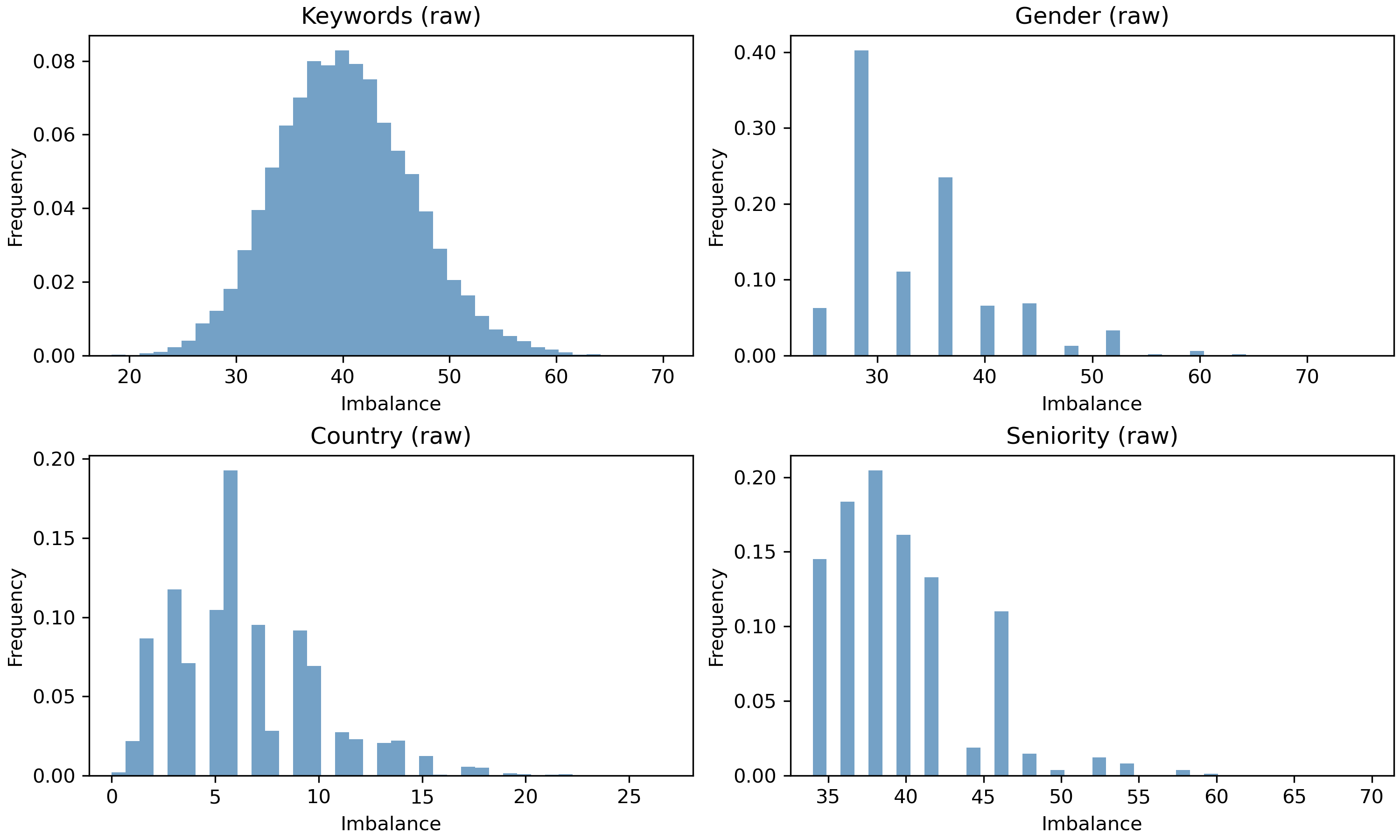}
\end{tabular}
\end{center}
\caption{\label{fig:raw_distrib} 
Imbalance distributions for a simulated population of 30,000 individuals with the properties shown in Figure~\ref{fig:reviewers}.}
\end{figure} 

\subsection{Keyword imbalance}
\label{sec:keywords}

This function computes a fitness score based on keyword imbalance across panels. The goal is to penalize panels where certain keywords are overrepresented. The imbalance is calculated by squaring the counts of each keyword in a panel, which amplifies the penalty for repeated keywords. The choice of using a quadratic sum is somewhat arbitrary, and it is meant to amplify the effect of repeated keywords, ensuring that panels with multiple overlapping keywords are penalized disproportionately more than those with only a few minor overlaps. A similar approach is adopted for the other imbalances described in the next subsections.
Let $i$ indicate the $i$-th member of panel $p$, with $1\leq i \leq n$ and $1\leq p \leq M$. Let then $K_{i,p}$ be the set of keywords specified by the $i$-th reviewer in panel $p$.
We then define the list $K_p$ of non-unique keywords specified by the members of the generic panel $p$ as the following union:

\[
K_p = \bigcup_{i=1}^{n} K_{i,p}
\]

We then compute the number of times a certain keyword $k$ occurs in each panel $p$ as:

\[
c_{k,p}=count(k,K_p)
\]

The imbalance score is computed by summing the squares of counts $c_{k,p}$ for all keywords $k$ in each panel $p$, but only for $c_{k,p}>1$ (keywords appearing more than once):

\[
\text{Keyword Imbalance} =
\sum_{p=1}^{M} \sum_{k \in K_p}
\begin{cases}
c_{k,p}^{2}, & \text{if } c_{k,p} > 1 \\[6pt]
0, & \text{otherwise.}
\end{cases}
\]

The FoM can be refined by introducing positional weights to account for the relative importance of each reviewer’s keywords. Let $K_{i,p}=[k_{i,p}^1,k_{i,p}^2,\dots,k_{i,p}^{n_{k,i,p}}]$ be the ordered list of keywords provided by reviewer $i$ in panel $p$, where each reviewer may specify a different number of keywords $n_{k,i,p}$. Each position $j$ in the list (with $1\leq j\leq n_{k,i,p}$) is associated with a positional weight $w_j$, such that $w_1\geq w_2\geq\ldots\geq w_{n_{max}}$, where $n_{max}$ is the maximum number of keywords. For instance, if $n_{max}=5$, one could have $w=[1.0,0.8,0.6,0.4,0.4]$. For each keyword $k$, we define its total weighted presence in panel $p$ as:

\[
W_{k,p} = \sum_{i=1}^{n} \sum_{j=1}^{n_{k,i,p}} 
w_j \, \mathbf{1}_{\{\,k_{i,p}^{j} = k\,\}}
\]

where $\mathbf{1}_{\{\cdot\}}$ is the indicator function:

\[
\mathbf{1}_{\{\,k_{i,p}^{j} = k\,\}} =
\begin{cases}
1, & \text{if } k_{i,p}^{j} = k \\[6pt]
0, & \text{otherwise.}
\end{cases}
\]

We also define $m_{k,p}$ as the number of distinct reviewers in panel $p$ who mention keyword $k$:

\[
m_{k,p} = \sum_{i=1}^{n} \mathbf{1}_{\{\,k \in K_{i,p}\,\}}
\]

The keyword imbalance for the entire set of panels is then:

\[
\text{Keyword Imbalance} =
\sum_{p=1}^{M} \sum_{k \in K_p}
\begin{cases}
W_{k,p}^{2}, & \text{if } m_{k,p} > 1 \\[6pt]
0, & \text{otherwise.}
\end{cases}
\]

That is, a keyword contributes to the imbalance only if it appears in more than one reviewer’s list within the same panel, and its contribution grows quadratically with the total sum of its positional weights across those reviewers. A low score indicates that keywords are well distributed, with minimal overlap among panel members. On the contrary, a high score implies that the panel contains multiple reviewers sharing highly weighted keywords, indicating topical redundancy or lack of diversity. This penalizes panels with a larger occurrence of keywords with higher weights, which in turn means higher concentration of reviewers with the same expertise.

\subsection{Gender Imbalance}
\label{sec:gender}

The gender imbalance score quantifies how far each panel’s gender composition deviates from an equal distribution across the three categories F (female), M (male), and N (not specified). In the ESO reviewer database, N is an explicit category selected by reviewers who prefer not to declare their gender.

Let \( p = 1, 2, \ldots, M \) index panels. 
For panel \( p \), let \( n_{p,F} \), \( n_{p,M} \), and \( n_{p,N} \) be the number of reviewers 
in categories F, M, and N, respectively, with 
\( n_{p,F} + n_{p,M} + n_{p,N} = n \). 
The ideal per-category share is \( n/3 \).\footnote{The equal distribution adopted here is only an illustrative example. In practice, the target distribution could instead be defined by the overall gender composition of the reviewer pool.}
The gender imbalance is computed as follows:

\[
\text{Gender Imbalance} =
\sum_{p=1}^{M} \sum_{g \in \{F, M, N\}}
\left( n_{p,g} - \frac{n}{3} \right)^{2}
\]

For \( n = 6 \), the minimum is \( 0 \) at a perfectly balanced panel \( (2,2,2)\), 
and the maximum is \( 24 \) at \( (6,0,0) \) (or any permutation).

\subsection{Country Imbalance}
\label{sec:country}

The country imbalance quantifies the lack of geographic diversity within each panel. 
It is computed from the squared difference between the total number of panel members 
and the number of unique countries represented. 
Panels with fewer unique country affiliations are therefore penalized more heavily. 

Let \( u_p \) be the number of distinct countries represented in panel \( p \). 
The overall country imbalance is given by:

\[
\text{Country Imbalance} =
\sum_{p=1}^{M} (n - u_p)^2
\]

For a panel with six members, the maximum imbalance (\(25\)) occurs when all members 
are affiliated with the same country, while the minimum imbalance (\(0\)) occurs 
when all six members come from different countries.

\subsection{Seniority Imbalance}
\label{sec:seniority}

This FoM evaluates the distribution of professional seniority levels across panels 
and penalizes those panels that exhibit an uneven composition. 
The objective is to promote a balanced representation of members at different career stages. 

For the purpose of this formulation, we consider three seniority levels: 
Junior (J), Mid-level (M), and Senior (S). 
Let \( n_{l,p} \) be the number of members with seniority level \( l \) in panel \( p \), 
where \( l \in \{J, M, S\} \). 
The seniority imbalance is defined as:

\[
\text{Seniority Imbalance} =
\sum_{p=1}^{M} \sum_{l \in \{J, M, S\}}
\left( n_{l,p} - \frac{n}{3} \right)^{2}
\]

\begin{figure} [t]
\begin{center}
\begin{tabular}{c}
\includegraphics[width=\linewidth]{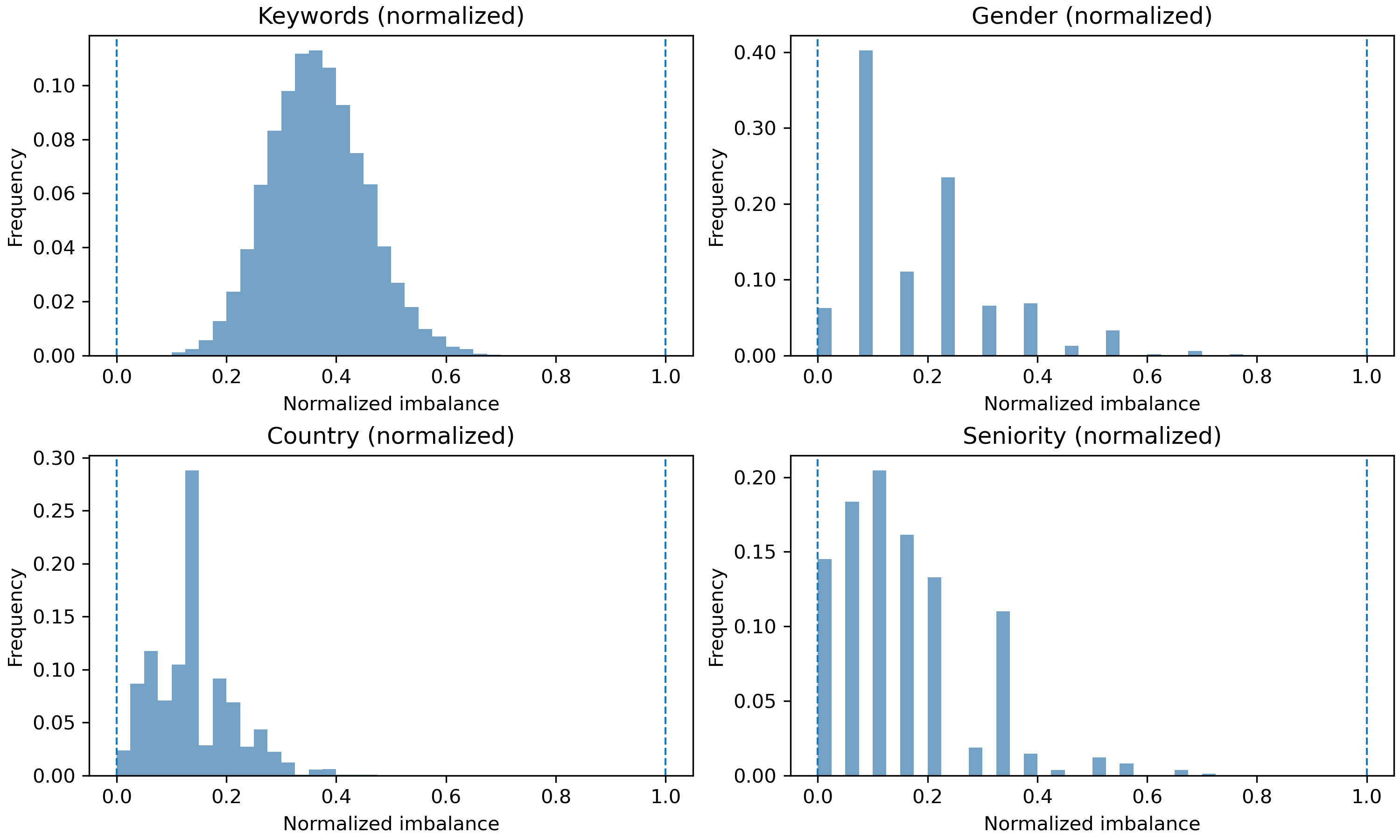}
\end{tabular}
\end{center}
\caption{\label{fig:norm_distrib} 
Normalised Imbalance distributions for the population presented in Figure~\ref{fig:raw_distrib}.}
\end{figure}

\subsection{Fitness Components Normalisation}
\label{sec:normalisation}

The various components of the fitness function can exhibit different numerical ranges, depending on their respective definitions. To ensure that they are appropriately weighted when combined into a single objective (see next section), each component must first be normalized to a representative scale.

Given a fixed pool of reviewers and a set of hard constraints (panel sizes, fixed chairs, etc.), the exact minimum and maximum of an imbalance measure can in principle be determined through a dedicated optimization procedure. However, our goal is not to certify absolute extrema, but rather to obtain representative lower and upper anchors for normalization. For this purpose, a randomized approach is simpler, computationally efficient, and sufficiently accurate.

The procedure consists of generating a large population of valid random panel assignments, computing the corresponding imbalance values, and identifying candidate minima and maxima. These solutions are then refined through a simple local search based on random reviewer swaps that preserve all constraints. The resulting extrema, denoted by $I_1$ and $I_2$, are used as normalization anchors according to the following scheme:

\begin{enumerate}
\item Generate a large batch of valid random assignments;
\item For each FoM, identify the best and worst individuals;
\item Apply a local hill-climbing refinement to further decrease or increase the corresponding FoM;
\item Use the resulting minimum and maximum values as normalization anchors $I_1$ and $I_2$.\footnote{A simpler alternative is to adopt the minimum and the 95th percentile of the sampled distribution. Tests on real reviewer pools indicate that this approximation yields very similar results.}
\end{enumerate}

Once these values are computed, the normalization is achieved using this simple linear transformation:

\[
\text{Normalised Imbalance} =
\frac{\text{Imbalance}-I_1}{I_2-I_1}
\]

To illustrate this procedure, we consider the real case of the ESO review panels for Period 117, D category. This configuration includes 24 experts distributed across four panels. The distributions of their properties are shown in Figure~\ref{fig:reviewers}. This example represents a typical scenario in which reviewers come from around a dozen different countries, display an overall well-balanced gender composition, and are predominantly senior scientists. Their expertise covers roughly 50 distinct keywords, some of which are shared by multiple reviewers.
A large population of candidate panel compositions can then be generated, each reflecting the same global characteristics as the real sample. 
The extrema obtained through the sampling and refinement procedure define the normalization anchors \(I_1\) and \(I_2\). The corresponding distributions of the four imbalance measures are shown in Figure~\ref{fig:raw_distrib}.

These distributions illustrate the typical ranges and shapes of the four Figures of Merit (FoMs) and emphasize the need for proper normalization. Notably, the keyword imbalance exhibits significantly higher values and a broader spread compared to the other FoMs, which show lower minima and narrower peaks.
The effect of the normalization transformation described above on the distributions in Figure~\ref{fig:raw_distrib} is shown in Figure~\ref{fig:norm_distrib}. After normalization, the four FoMs exhibit comparable ranges, allowing them to be directly combined into a single, weighted measure of total imbalance.

\subsection{Weighted Total Imbalance}

The overall imbalance is obtained as a weighted combination of the individual imbalance components, 
where the weights reflect the relative importance assigned to each aspect of panel composition:

\[
\text{Total Imbalance} =
w_k I_{\text{keywords}} +
w_g I_{\text{gender}} +
w_c I_{\text{country}} +
w_s I_{\text{seniority}}
\]

The weights are constrained by the normalization condition:

\[
\sum_{j \in \{k, g, c, s\}} w_j = 1
\]

By adjusting these weights, one can emphasize specific aspects of balance depending on 
the objectives of the selection process, for instance, assigning higher weight to gender balance 
to promote inclusivity, or to keyword balance to ensure adequate coverage of expertise areas.

\begin{figure} [t]
\begin{center}
\begin{tabular}{c}
\includegraphics[width=\linewidth]{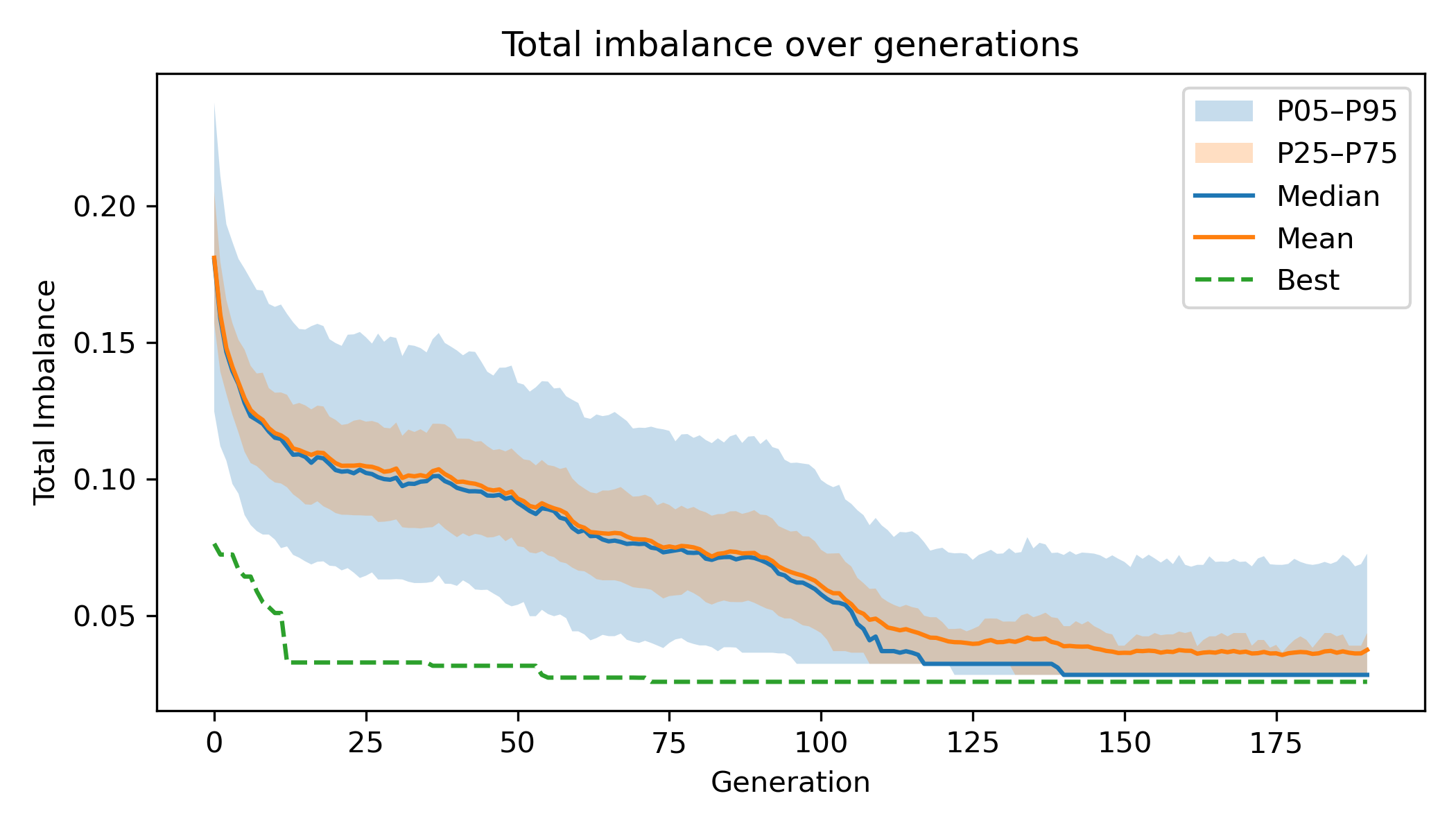}
\end{tabular}
\end{center}
\caption{\label{fig:evolution} 
Evolution of the total imbalance across generations. Each generation includes 3000 different panel compositions. The shaded areas indicate the 25th-75th (P25-P75) and 5th-95th (P05-P95) percentile intervals. The dashed line traces the evolution of the best individual of the given generation.}
\end{figure}

\section{CONVERGENCE CRITERION}
\label{sec:convergence}

The genetic algorithm is terminated using a stagnation-based convergence criterion. Rather than fixing the number of generations a priori, the evolutionary process is allowed to continue until no significant improvement in the fitness of the best solution is observed over a predefined number of consecutive generations, defined as the stagnation limit. The rationale behind this approach is that, during the initial phases of the search, crossover and mutation operators typically produce frequent improvements in the population fitness. As the population evolves and approaches a near-optimal solution, improvements become progressively smaller and eventually cease. At this stage, additional generations contribute little to the quality of the solution while increasing computational cost.

Stagnation is monitored by tracking the fitness value of a reference individual in each generation. This can be the best, but also a more robust one, like the median individual. A counter is incremented whenever the best fitness changes by less than a predefined tolerance between successive generations. When the number of consecutive stagnant generations exceeds a specified threshold, the algorithm is considered to have converged and the evolutionary process is terminated.

This criterion offers several advantages over a fixed-generation stopping rule. First, it adapts the computational effort to the complexity of the optimization problem, allowing difficult instances to evolve for longer while terminating early when convergence is reached rapidly. Second, it prevents unnecessary evaluations once the search has effectively exhausted the available improvements. Finally, it provides a practical indication that the population has reached a stable region of the solution space where further exploration is unlikely to yield substantially better panel compositions.

\section{TESTS AND RESULTS}
\label{sec:results}

The procedure discussed in the above sections has been implemented using the DEAP Python module.\cite{DEAP} Following the standard genetic algorithm framework described by Goldberg,\cite{Goldberg89} the optimization employs selection, elitism, mutation, and crossover operators.

The method was subsequently tested using the real data of the 78 reviewers recruited for ESO Period 117. In the following, I illustrate the case of Category D (Stellar Evolution), which consists of four panels with six members each (see Figure~\ref{fig:reviewers} for the overall distribution of reviewer properties). The evolution was run with the following parameters: population size = 3000 individuals, mutation probability = 0.5, crossover probability = 0.5, maximum number of generations = 200. 

The relative weights on the four imbalances were set to $w_k$=1.0, $w_g$=0.9, $w_c$=0.8, $w_s$=0.5 (corresponding to normalized weights 0.32, 0.28, 0.25, 0.15). The distribution of the overall imbalance, and its typical statistical indicators (best, mean, median, inter-quartile range, ...), can be monitored across the various generations. For the sake of illustration, no stop criterion was used and the algorithm was run for 200 generations. The results are presented in Figure~\ref{fig:evolution}.

Several interesting features emerge from inspection of this figure. 

\begin{enumerate}
\item The best individual achieves a relatively low total imbalance already within the initial population and improves rapidly during the first few generations. Thereafter, only marginal improvements are observed, indicating that near-optimal solutions are identified early in the evolutionary process.

\item The overall population imbalance, represented by the mean and median values, decreases more gradually. The continued reduction over many generations indicates that the evolutionary operators progressively improve not only the best solution but also the population as a whole. The rate of improvement eventually slows, approaching convergence after approximately 100 generations.

\item The spread of imbalance values within the population, as measured by both the interquartile range (P25--P75) and the broader P05--P95 interval, decreases steadily throughout the simulation. This reduction reflects a progressive homogenization of the population as individuals converge towards increasingly similar levels of fitness.
\end{enumerate}

These results highlight an important aspect of the genetic algorithm. The outcome of the optimization is not merely a single high-quality solution, but rather a final population composed predominantly of individuals that are substantially superior, according to the fitness function, to those present in the initial random population.

This observation has practical implications for panel composition. Given the characteristics of the reviewer pool, it is likely that numerous panel configurations exhibit very similar values of total imbalance. Consequently, the final population can be ranked according to fitness, after which additional criteria not explicitly included in the optimization function may be applied. Such criteria can be used to discriminate among solutions of comparable quality without requiring modifications to the genetic algorithm itself.

A simple example concerns the geographical distribution of panel members. In many contemporary review processes, meetings are conducted virtually and must accommodate participants located in widely separated time zones. Configurations that simultaneously include reviewers from regions such as Australia and North America may therefore be operationally undesirable, despite exhibiting excellent fitness according to the imbalance metric.

The most straightforward strategy is to rank the individuals in the final population by increasing total imbalance and then select the highest-ranked solution that satisfies the additional operational constraint. This approach preserves the advantages of the genetic optimization while retaining flexibility to incorporate practical considerations that are difficult to encode directly within the fitness function.

\section{Conclusions}
\label{sec:conclusions}

I have presented a genetic algorithm designed to optimize the composition of scientific review panels under a set of practical constraints commonly encountered in peer-review systems. The method represents panel assignments through a chromosome-based encoding that preserves fixed panel sizes and accommodates additional constraints such as pre-assigned panel chairs. The quality of a panel configuration is evaluated through a fitness function based on four imbalance indicators: scientific expertise, gender, country affiliation, and professional seniority. By normalizing these indicators and combining them through adjustable weights, the algorithm can be tailored to the specific priorities of a given review process.

Tests performed using real reviewer data from the ESO proposal evaluation process demonstrate that the genetic algorithm rapidly identifies panel configurations with significantly lower imbalance than those found in the initial random population. While the best solutions emerge after relatively few generations, the continued evolution of the population progressively improves the average quality of candidate configurations and reduces the spread of fitness values, leading to a population composed predominantly of high-quality solutions.

An important outcome of this approach is that the optimization does not merely produce a single preferred panel composition. Instead, it generates a diverse set of solutions with comparable fitness, providing additional flexibility for the final selection. Operational constraints that are difficult to encode directly in the fitness function, such as time-zone compatibility, conflict management, or other logistical considerations, can subsequently be applied to the final ranked population without requiring modifications to the optimization procedure.

Although developed for the ESO review process, the methodology is general and can be readily adapted to other peer-review systems that rely on panel-based evaluation. More broadly, the approach is applicable to any assignment problem in which individuals must be distributed among groups while simultaneously balancing expertise, diversity, and organizational constraints.

The implementation presented in this paper adopts the keyword system currently used at ESO, in which reviewers self-identify their areas of expertise through a set of scientific keywords. This approach has the advantage of being simple, transparent, and directly controlled by the reviewers themselves. However, the optimization algorithm is largely agnostic to the specific mechanism used to characterize expertise. In principle, the keyword vectors could be replaced by more sophisticated representations derived from bibliometric information, such as publication records, citation networks, topic modelling, or other natural-language processing techniques applied to the scientific literature authored by the reviewers. Such approaches may provide a more objective and continuously updated description of scientific expertise, although they also introduce additional complexity and potential biases related to publication databases and profiling methodologies. A practical implementation of publication-based expertise characterization has been presented by Kerzendorf et al.\cite{Kerzendorf2020} and Strolger et al.\cite{Strolger23}, who derived reviewer topical profiles from the NASA Astrophysics Data System bibliographic records and used them for reviewer matching. More recently, Carpenter at al.\cite{Carpenter2025} developed an alternative approach for the ALMA proposal review process, in which expertise profiles are inferred from the topics of previously submitted proposals rather than from publication records.

A detailed comparison between self-declared and automatically derived expertise indicators is beyond the scope of this work. Nevertheless, the framework described here is sufficiently general that alternative expertise representations could be incorporated with only minor modifications to the fitness function, preserving the overall optimization strategy.

\acknowledgments 
 
I am grateful to F. Primas, A. De Cia, F. Sogni and D. Dorigo for fruitful discussions on the implementation of this algorithm in the ESO proposal handling process. 

\bibliography{report} 
\bibliographystyle{spiebib} 

\end{document}